\documentclass[a4paper,11pt]{article}
\pdfoutput=1 

\usepackage{jheppub} 

\usepackage[T1]{fontenc} 

\title{\boldmath  Vector and Scalar Mesons' Mixing from QCD Sum Rules}

\author[a]{Ze-Sheng Chen,}
\author[b]{Zhu-Feng Zhang,}
\author[c]{Zhuo-Ran Huang,}
\author[d]{T. G. Steele,}
\author[a]{and Hong-Ying Jin}

\affiliation[a]{Institute of Modern Physics, Department of Physics, Zhejiang University, Hangzhou, 310027,China}
\affiliation[b]{Department of Physics, Ningbo University, Ningbo, 315211, China}
\affiliation[c]{Institute of High Energy Physics, Chinese Academy of Sciences, Beijing, 100049, China}
\affiliation[d]{Department of Physics and Engineering Physics, University of Saskatchewan, Saskatoon, Saskatchewan, S7N 5E2, Canada}
\emailAdd{ventuschen@zju.edu.cn}
\emailAdd{zhangzhufeng@nbu.edu.cn}
\emailAdd{huangzr@ihep.ac.cn}
\emailAdd{tom.steele@usask.ca}
\emailAdd{jinhongying@zju.edu.cn}

\abstract{We study $\bar qq$-hybrid mixing for the light vector mesons and
 $\bar qq$-glueball mixing for the light scalar mesons in Monte-Carlo based
QCD Laplace sum rules. By calculating the two-point correlation function of a
vector $\bar q\gamma_\mu q$ (scalar $\bar q q$) current and a hybrid (glueball) current we
are able to estimate the mass and the decay constants of the corresponding
mixed ``physical state'' that couples to both currents. Our results
do not support strong quark/gluonic mixing for either  the $1^{--}$ or  the $0^{++}$ states.}

\begin{document}
\maketitle
\flushbottom

\section{Introduction}

\noindent
 Many experimental observations of new hadrons have suggested more abundant meson
 spectroscopy than what is suggested by the quark model \cite{ref_article1},
 and  many other models have been constructed to explain meson states  including
 hybrids, tetraquarks, glueballs, etc \cite{ref_article3,ref_article5,ref_article6,Liu:2019zoy,Chen:2016qju}. By virtue of QCD  sum rules (QCDSR) based on a QCD correlation function plus an
 appropriate spectral density, one can study the constituents of hadrons by
 using different interpolating currents \cite{ref_article4}. Fruitful results on exotic states
 including heavy and light multiquark states, hybrids and glueballs have been obtained \cite{ref_article5,ref_article6,ref_article2,Matheus:2006xi,ref_article7,ref_article8,ref_article19,ref_article16,ref_article20,Huang:2016upt,Huang:2016rro,Fu:2018ngx,Zhang:2011qza,Chen:2013zia,Chen:2013pya,Chen:2008qw,Jiao:2009ra,Albuquerque:2009ak,Zhang:2011jja,Pimikov:2017bkk,Qiao:2015iea,Harnett:2008cw,Harnett:2000fy,Ho:2016owu,Chen:2014fza,Narison:1996fm,Narison:2005wc}. 
 However, we know these states and the ordinary $\bar qq$ mesons can mix with each other via
 QCD interactions. The mixing scenario may affect the analyses of QCDSR based on the pure
 constituent scenario.  Because of non-perturbative QCD, it is not easy to understand hadron mixing quantitatively. Some researchers used a Low Energy Theorem or other
 methods \cite{ref_article25,ref_article26} to study the constituents of possible mixed
 states with different conclusions.
  In this work, we 
  build upon previous  QCDSR-based approaches \cite{Narison:1984bv,Harnett:2008cw,Palameta:2018yce,Palameta:2017ols} to deal with this problem.

In QCDSR, one normally calculates the two-point correlator of a current
and its Hermitian conjugate, and by inserting a complete set of particle
eigenstates between the two currents, one can pick up a state which has the
``strongest signal'' in the spectral density, i.e., the state with a relatively
low mass and relatively strong coupling to the current. In the scenario of  leading-order perturbation theory, the structure of the current reflects the dominant
constituents of the corresponding state. However, because of non-perturbative and higher-loop QCD effects, it is possible that the  state also couples to different
currents with comparable strength, therefore it is interesting to consider the
two-point correlator of two different currents (see e.g., Refs.~\cite{Narison:1984bv,Harnett:2008cw,Palameta:2018yce,Palameta:2017ols}). One can still insert a complete
set of eigenstates and use a Borel transform to pick up the state with the ``strongest signal''.
Certainly, such a state should have a relatively low mass and couple to both currents
relatively strongly. If such a state involves strong mixing, which means it contains
large constituent of both pure states, the corresponding couplings to both currents can have considerable values. Therefore the ``mixing strength'' can be reflected in the product of the decay constants, for which we will give a more precise definition
in the next section. By estimating the mass, the ``mixing strength'' and taking into
account experimental results, one can get insight into the constituent composition of
the corresponding states.

In this paper, we use  Monte-Carlo based Laplace QCDSR methods for non-diagonal correlation functions to study two quantum numbers $1^{--}$ and $0^{++}$ that have
long been considered to involve the meson mixing of light-quark $\bar q q$ and gluonic (glueball/hybrid) constituents \cite{ref_article1}. 

For the $1^{--}$ channel, there are quite a few vector states found in experiments \cite{ref_article1}
which are in principle difficult to be all explained by the naive quark model (e.g., the  $\rho(1450)$ and $\rho(1570)$ are too close to each other,
which violates the rule of Regge trajectories). Because the $1^{--}$ hybrid
is expected to be degenerate with the $1^{-+}$ hybrid \cite{Barnes:1982tx,ref_article3}
which is believed to be around 2 GeV \cite{ref_article20,ref_article19}, it is therefore interesting to see whether there is a large mixing between $1^{--}$ $\bar qq$ states
and hybrid states. Although Laplace QCDSR analyses have  been applied to $\bar q q$-hybrid mixing in heavy-quark $1^{--}$ systems \cite{Palameta:2017ols}, the corresponding light-quark systems have not  previously been studied.

For the $0^{++}$ sector, many states lie in the range 0.6--1.7 GeV, most of which have not been well understood. It is
generally believed that some of them can be glueball candidates \cite{ref_article1}.
The $0^{++}$ glueball mass is predicted to be 1.5--1.7 GeV in Lattice QCD, and large
mixing between the $\bar qq$ and the glueball is generally expected \cite{ref_article30}.
Investigation of the mixing between the $0^{++}$ $\bar qq$ states and the glueballs can therefore contribute to the interpretation of the scalar mesons. 
This work extends and is complementary to a previous QCDSR analysis of the $0^{++}$ mixed $\bar q q$-gluonic correlator \cite{Harnett:2008cw} in a few significant ways. First, Laplace sum-rules are used in contrast to the Gaussian sum-rule analysis of  Ref.~\cite{Harnett:2008cw}, thereby exploring a fundamentally different weighting of QCD perturbative and non-perturbative  contributions.  Second, the effect of higher-dimension condensates and  higher-loop quark condensate effects are considered. Finally, a different quantification of the mixing degree  is developed and analyzed.

Although our primary interest in the $0^{++}$ sector is the mixture of $\bar q q$ and glueball components, our analysis does not exclude or constrain a multi-quark scenario for the scalar mesons. There is a vast literature that encompasses different aspects of the $\bar q q$ and $\bar q q \bar q q$ components of the scalar mesons, with methodologies ranging from chiral Lagrangians to QCD sum-rules (see e.g., Refs.~\cite{Dai:2019lmj,Mec,close,mixing,NR04,06_F,08_tHooft,global,07_FJS4,05_FJS,Jaf,Zhang:2000db,Brito:2004tv,Chen:2007xr}). An inverted mass spectrum, as first noted in the MIT bag model \cite{Jaf}, is an important aspect of the $\bar q q\bar q q$ scenario.

Our methodology is introduced in section~\ref{sec:II}. Then we discuss $1^{--}$ states and $0^{++}$ states in section~\ref{sec:III}
and section~\ref{sec:IV} respectively. Finally we give our summary and conclusions in
the last section.

\section{Fitting method and mixing strength}\label{sec:II}

In QCDSR, the hadronic  mixing normally is studied with the two-point correlator
\begin{equation} \label{mix}
\begin{aligned}
\Pi\left ( q^{2} \right )&=i\int d^{4} e^{iqx}\left \langle 0 \left |T\left \{(j_{a}\left ( x \right )+c j_b(x) )(j_{a}^{+}( 0)+ c j_b^{+}(0) )\right \}\right| 0 \right \rangle\\
&=\Pi_a(q^2)+c\Pi_{ab}(q^2)+c^2\Pi_b(q^2),
\end{aligned}
\end{equation}
where $j_a$ and $j_b$ have the same quantum number, $c$ is a real parameter to describe the mixing strength, and
\begin{equation}
\begin{aligned}
\Pi_a\left ( q^{2} \right )&=i\int d^{4} e^{iqx}\left \langle 0 \left |T(j_{a}(x)j_a^{+}(0))\right| 0 \right \rangle,\\
\Pi_b\left ( q^{2} \right )&=i\int d^{4} e^{iqx}\left \langle 0 \left |T(j_{b}(x)j_b^{+}(0))\right| 0 \right \rangle,\\
\Pi_{ab}\left ( q^{2} \right )&=i\int d^{4} e^{iqx}\left \langle 0 \left |T(j_{a}(x)j_b^{+}(0)+j_{b}(x)j_a^{+}(0))\right| 0 \right \rangle.
\end{aligned}
\end{equation}
The correlator obeys a dispersion relation
\begin{equation} \label{disper1} 
\begin{aligned}
\Pi\left ( q^{2} \right )&=\Pi_a(q^2)+c\Pi_{ab}(q^2)+c^2\Pi_b(q^2)\\&=\frac{(q^2)^N}{\pi }\int_{0}^{\infty }ds\frac{\textrm{Im}\Pi_a( s)+c\textrm{Im}\Pi_{ab}( s)+c^2\textrm{Im}\Pi_b( s)}{s^N(s-q^{2}-i\epsilon)}+\ldots~\textrm{,}
\end{aligned}
\end{equation}
where $N$ is positive integer 
that  depends on the dimension of the corresponding current, and dots on the right hand side represent polynomial subtraction terms to render $\Pi(q^2)$ finite.

Obviously (\ref{disper1}) can be divided into three independent equations
\begin{equation}
\begin{aligned}
\Pi_a(q^2)&=\frac{(q^2)^N}{\pi }\int_{0}^{\infty }ds\frac{\textrm{Im}\Pi_a( s)}{s^N(s-q^{2}-i\epsilon)}+\ldots~\textrm{,}\\ 
\Pi_{ab}(q^2)&=\frac{(q^2)^N}{\pi }\int_{0}^{\infty }ds\frac{\textrm{Im}\Pi_{ab}( s)}{s^N(s-q^{2}-i\epsilon)}+\ldots~\textrm{,}\\
\Pi_b(q^2)&=\frac{(q^2)^N}{\pi }\int_{0}^{\infty }ds\frac{\textrm{Im}\Pi_b( s)}{s^N(s-q^{2}-i\epsilon)}+\ldots~\textrm{.}
\end{aligned}
\end{equation}
By tuning the parameter $c$, one can obtain the best sum rules via Eq.~(\ref{disper1}).
However, the mixing is often a small effect  and is easily obscured by the dominant constituent in the full correlator $\Pi(q^2)$.
In order to highlight the information from the mixing, it is better to consider the mixing correlator $\Pi_{ab}(q^2)$ alone.
In the following, we will consider the correlator $\Pi_{ab}(q^2)$ with $j_{a\mu}=\epsilon _{\mu \phi \alpha \beta }
\bar{q}(x)gG_{\alpha \phi }\gamma_{5}\gamma _{\beta }q(x)$ and $j_{b\nu}=\bar{q}(x)\gamma _{\nu}q(x)$ for $1^{--}$,
and $j_{a}=m\bar{q}(x)q(x)$ and $j_{b}=2$Tr$(\alpha_{s}G_{\mu\nu}G_{\mu\nu})$ for $0^{++}$.

The correlator $\Pi_{ab}(q^2)$ can be calculated using the operator product expansion (OPE) \cite{ref_article4}.
On the other hand, by using the narrow resonance spectral density model, i.e.,
\begin{equation}
\textrm{Im}\Pi^{\textrm{(phen)}}_{ab} \left ( s \right )=\pi \delta \left
( s-m_{0}^{2} \right )(f_{a}f_{b}^{*}+f_{b}f_{a}^{*})+\theta \left ( s-s_{0} \right )\textrm{Im}\Pi_{ab} ^{(\textrm{OPE})}\left ( s \right )\textrm{,}
\end{equation}
where $f_{a}$ and $f_{b}$ are the  respective couplings of the ground state to the corresponding currents,
and $s_0$ is the continuum threshold which separates the contributions from excited states, we also can express the correlator $\Pi_{ab}(q^2)$ through
the dispersion relation:
\begin{equation}
\Pi^{\textrm{(phen)}}_{ab}(q^2)=\frac{(q^2)^N}{\pi }\int_{0}^{\infty }ds\frac{\textrm{Im}\Pi^{\textrm{(phen)}}_{ab}( s)}{s^N(s-q^{2}-i\epsilon)}+\ldots~.
\end{equation}
By demanding equivalence of the phenomenological and OPE expressions, we obtain the master equation in for QCDSR:
\begin{equation}
\Pi_{ab}^{\textrm{(OPE)}}=\frac{(q^2)^N}{\pi }\int_{0}^{\infty }ds\frac{\textrm{Im}\Pi^{\textrm{(phen)}}_{ab}( s)}{s^N(s-q^{2}-i\epsilon)}+\ldots~.
\label{disper}
\end{equation}

After applying the Borel transformation operator $\hat B$ to \eqref{disper},  the subtraction terms are eliminated and the master equation can be written as
\begin{equation}
\begin{split}
&R^{\textrm{(OPE)}}(\tau)\equiv\frac{1}{\tau}\hat{B}\Pi^{\textrm{(OPE)}}\left (q^{2} \right )=\frac{1}{\pi }\int_{0}^{\infty }ds\,
\textrm{Im}\Pi^{\textrm{(OPE)}} \left ( s \right )e^{-s\tau}\\
=&R^{\textrm{(phen)}}(\tau)\equiv \frac{1}{\pi } \left [\pi  (f_{a}f_{b}^{*}+f_{b}f_{a}^{*})e^{-m^{2}_{0}\tau}+\int_{s_{0}}^{\infty} ds \,\textrm{Im}\Pi ^{\textrm{(OPE)}}\left ( s \right )e^{-s\tau}\right ].
\end{split}
\end{equation}
By placing the contributions from excited states on the OPE side, we finally obtain
\begin{equation} \label{2} 
\int_{0}^{s_{0} }ds \,\textrm{Im}\Pi^{\textrm{(OPE)}} \left ( s \right )e^{-s\tau}=(f_{a} f_{b}^{*}+f_{a}^* f_{b})e^{-m^{2}_{0}\tau}\textrm{,}
\end{equation}
where $m_{0}$ is mass of the   state which has the  strongest signal.  Because each particle's
contribution is partitioned into exponential function and the contribution of excited states would be
quickly suppressed by $e^{-m_{}^{2}\tau}$, $m_{0}$ should not be much heavier than ground
state's mass. Meanwhile, the value of $ f_{a} f_{b}^{*}+f_{a}^* f_{b}$ plays an important role.
If  there is a state with a large mixing, its signal may overwhelm the ground state and be
selected out. Otherwise,  the ground state will dominate the correlator.  The master equation
(\ref{2}) is the foundation of our analysis.
Deviations from the narrow width approximation will be small provided that the $m_0\Gamma\tau\ll 1$ \cite{Elias:1998bq} which will be the case even for $f_0(500)$ widths   within the  $\tau$  range outlined below.

Because of the truncation of the OPE and the simplified assumption for the
phenomenological spectral density, Eq.(\ref{2}) is not valid for all values of $\tau$, thus the determination of the sum rule window, in which the validity of (\ref{2}) can be established, is very important.
In the literature, different methods are used in the determination of the $\tau$ sum rule window \cite{ref_article19,ref_article16}.
In this paper, we determine the range of $\tau$ by demanding that the resonance contribution is more than 50\% and the  highest dimension contribution (normally dimension-six, i.e., 6D) is less than 10\% in $R^{\textrm{(OPE)}}(\tau)$. However,  if higher dimension contributions are included (e.g., dimension-eight) in the analysis, we also require that the 8D contribution is  $\Lambda^2/\mu^2\approx1/4$ of 6D to ensure each higher dimensional portion has a reasonable distribution in the OPE series.

We use the Monte-Carlo based QCD sum rules analysis method to test values
for $s_{0}$, $m_0$, $f_{a} f_{b}^{*}+f_{a}^* f_{b}$ in order to find the best solution minimizing $\chi^2$ \cite{ref_article10}
\begin{equation}
\chi^2(s_{0},f_{a}f_{b}^{*}+f_{a}^* f_{b},m_{0})=\sum_{i=1}^{21} \left ( R ^{\textrm{(OPE)}} \left (\tau _{i}\right )-R ^{(\textrm{phen})}\left ( \tau _{i}\right )\right )^{2}/\sigma_{\textrm{OPE}}^{2}(\tau_{i})\textrm{,}
\end{equation}
where $\sigma_{\textrm{OPE}}(\tau_{i})$ is the standard deviation of $R ^{\textrm{OPE}} \left (\tau _{i}\right )$ at the point $\tau_i$,
and the sum rules window is divided into 20  equal intervals.

In order to estimate the mixing strength of the physical state strongly coupled to both the two different currents, we define
\begin{equation} \label{eq:mixing dgree}
\begin{aligned}
N\equiv&\frac{|f_{a} f_{b}^{*}+f_{a}^* f_{b}|}{2|f^{'}_{a}f^{'}_{b}|}\textrm{,}
\end{aligned}
\end{equation}
where $f^{'}_{a}$ and $f^{'}_{b}$ are decay constants of the relevant current with a pure state (i.e., the coupling that emerges in the diagonal correlation functions $\Pi_a$, $\Pi_b$).  
Eq.~\eqref{eq:mixing dgree} is analogous to the mixing parameter defined in Ref.~\cite{Hart:2006ps}.  By using appropriate factors of mass  in the definitions of $f^{'}_{a}$ and $f^{'}_{b}$, we
can therefore compare the magnitude of decay constants and estimate the mixing strength  self-consistently.
Obviously, a larger $N$ means stronger mixing strength of states. However, we cannot determine which part dominates the mixed state when $N$ is small. In this case, we compare the mass of the mixed
 state with the two relevant pure states, and we suggest that mixed state is dominated
 by the part whose pure state  mass  prediction is closest to the mixed state mass. The mixing strength depends on the definition and normalization of mixed state. For example, in Ref.~\cite{Narison:1984bv} the definition of the mixed state is
\begin{equation}
\left |\textrm{M}  \right \rangle=\cos \theta \left |\textrm{A}  \right \rangle+\sin \theta \left |\textrm{B}  \right \rangle ,
\end{equation}
where $\left |\textrm{M}  \right \rangle$ is a mixed state composed of pure states  $\left |\textrm{A} \right \rangle$  and  
$\left |\textrm{B} \right \rangle$
and $\theta$ is a mixing angle. In this definition and normalization of the mixed state,
we could see that $N\approx \cos \theta \sin \theta$, and $N\in \left ( 0, \frac{1}{2} \right )$.
Because of the different possible normalizations and mixed state definitions, we use  Eq.~\eqref{eq:mixing dgree} as a robust parameter to quantify mixing effects.


The central values of  the QCD input parameters are listed in Table \ref{tbl:bins}.
The input parameters including $\Lambda_{\text{QCD}}$, quark masses and  $m_{q}\left \langle \bar{q}q \right\rangle$ condensate are generated with 5\% uncertainties, and the others are generated with 10\% uncertainties, which is a typical uncertainty in QCDSR, allowing calculation of the $\chi^2$  fit to the two sides of
Eq.~(\ref{2}).  For the $s$ quark,
$\left \langle \bar{s}s\right \rangle$=0.8$\left \langle \bar{q}q\right \rangle$
will be used \cite{ref_article22}, and we set $\kappa=1.2$ when the four quark condensate contribution is included \cite{Shifman:1978by}. 
\begin{table}[ht]
\begin{center}
\caption{Phenomenological parameters (at the scale $\mu_{0}$=1 GeV).}
\label{tbl:bins} 
\begin{tabular}{ccc}
\hline
\hline
QCD Parameters & values & references\\
\hline
$m_{q}$/GeV            & 0.005 & \cite{ref_article1}\\$m_{s}$/GeV             & 0.126  & \cite{ref_article1}\\
$\Lambda_{\textrm{QCD}}$/GeV                                   &   0.343 & \cite{flag}\\
$m_{q}\left \langle \bar{q}q \right\rangle$          &   $-\frac{1}{4}f_{\pi}^{2}m_{\pi}^{2}$ & \cite{minireview}\\
$\left \langle g^{3}G^{3}\right \rangle$              &  8.2 GeV$^{2}$$ \left \langle\alpha_{s}G^{2}\right\rangle$  &
\cite{minireview}\\
$\alpha_{s}\left \langle\bar{q}q\right\rangle ^{2}$ /GeV$^{4}$   & $1.5\kappa \times 10^{-4}$  & \cite{minireview}\\
$\left \langle\alpha_{s}G^{2}\right\rangle$/GeV$^{4}$ &   0.07  & \cite{ref_article17,ref_article18}\\
$m_{q}\left \langle g\bar{q}Gq \right\rangle$             & 0.8 GeV$^{2}$ $m_{q}\left \langle\bar{q}q\right\rangle$ & \cite{ref_article22}\\

\hline
\end{tabular}
\end{center}
\end{table}

\section{Vector Hybrid and $\bar{q}q$ mixed state}\label{sec:III}

Both the hybrid current $j_{1\mu}=\epsilon _{\mu \phi \alpha \beta }
\bar{q}(x)gG^a_{\alpha \phi }T^a\gamma_{5}\gamma _{\beta }q(x)$
and the $\bar qq$ current $j_{2\nu}=\bar{q}(x)\gamma _{\nu }q(x)$ can
couple to $1^{--}$ states. To study the mixing of a $1^{--}$ state
which has hybrid and $\bar qq$ meson content, we start from the off-diagonal mixing
correlator described in the previous section, i.e.,
\begin{equation}
\Pi_{\mu\nu}(q^{2})=i\int d^{4}x \,e^{iq\cdot x}  \langle 0|
T(j_{1\mu}\left( x \right)j^{+}_{2\nu}(0)+j^{+}_{2\mu}\left( x \right )j_{1\nu}(0) ) |0 \rangle.
\end{equation}
Since $j_{2\nu}$ is conserved, $\Pi_{\mu\nu}(q^{2})$  has the form
\begin{equation}
\Pi_{\mu\nu}(q^{2})=\Pi_{\bar{q}Gq\bar{q}q}(q^{2})(q^{2}g_{\mu\nu}-q_{\mu}q_{\nu})\textrm{.}
\end{equation}


To calculate the OPE for $\Pi_{\bar{q}Gq\bar{q}q}(q^{2})$,
we use the massless quark propagator up to $O(q^{-5}$) \cite{ref_article23}
\begin{equation}
\begin{aligned}
S(q)=&S_{0}+\frac{ig}{2}G_{\mu\nu}S_{0}\partial_{\mu}\gamma_{\nu}S_{0}+\frac{g}{3}
D_{\alpha}G_{\mu\nu}S_{0}\partial_{\alpha}\partial_{\mu}\gamma_{\nu}S_{0} \\
&-\frac{ig}{8}D_{\beta}D_{\alpha}G_{\mu\nu}S_{0}\partial_{\beta}\partial_{\alpha}
\partial_{\mu}\gamma_{\nu}S_{0}-\frac{g^{2}}{4}G_{\rho\sigma}G_{\mu\nu}S_{0}
\partial_{\rho}\gamma_{\sigma}S_{0}\partial_{\mu}\gamma_{\nu}S_{0}~,
\end{aligned}
\end{equation}
where $S_{0}$ represents the free quark propagator,
$\partial_{\mu}=\partial/\partial q_{\mu}$ acts on all propagators to
the right, and $D_{\mu}=\partial_{\mu}-igA_{\mu}^{a}t^{a}$ acts only
on the nearest $G_{\mu\nu}$.

\begin{figure}[ht]
 \centering \includegraphics[width=0.8\columnwidth]{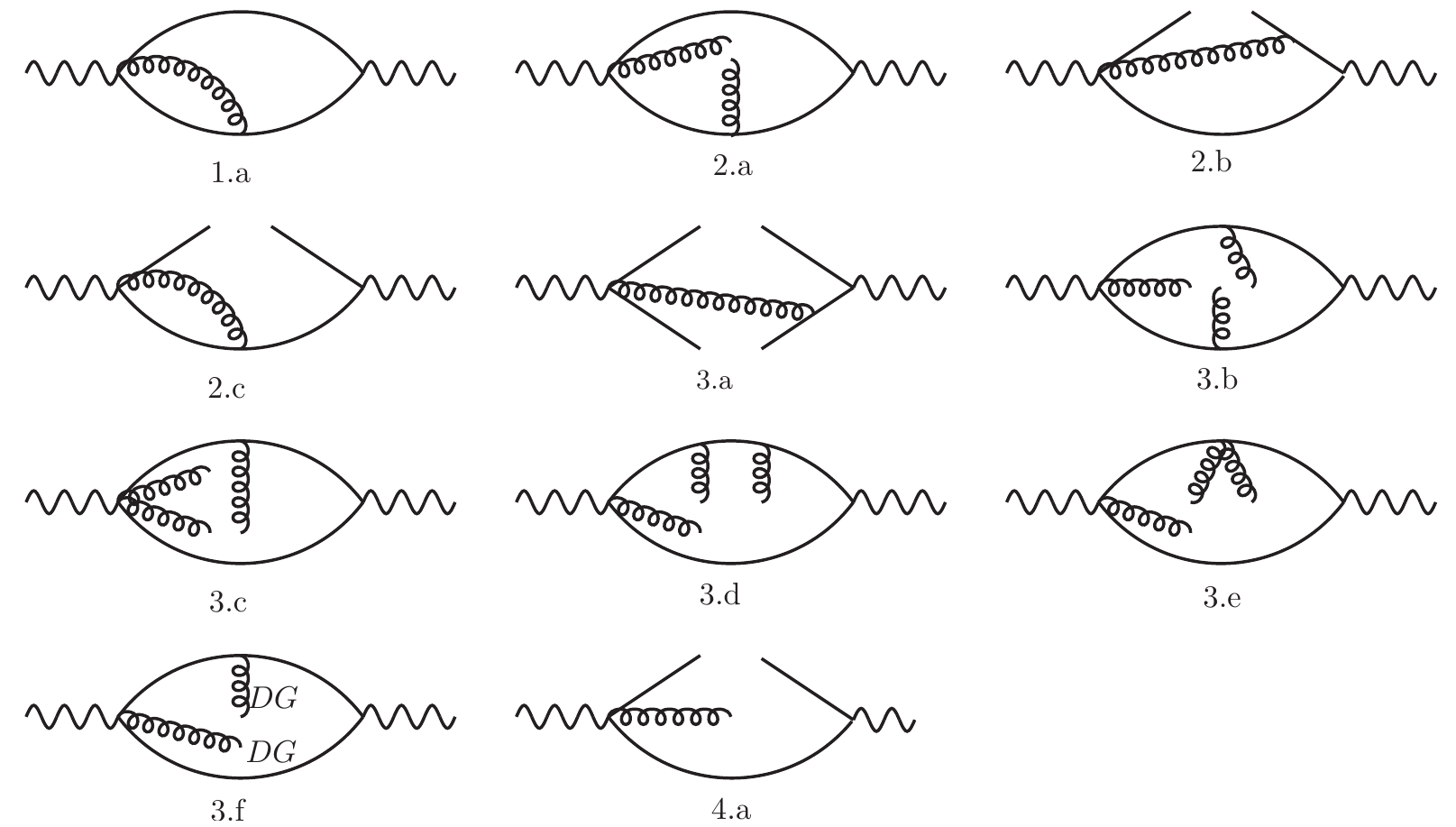}
  \caption{
                \label{fig1} 
              Feynman diagrams for the $\bar{q}q$-hybrid mixed state.
        }
\end{figure}

Collecting all the contributions to the correlator $\Pi_{\bar{q}Gq\bar{q}q}$ from Figure \ref{fig1}, we obtain
\begin{equation}
\begin{aligned}
\Pi_{\bar{q}Gq\bar{q}q}(q^{2})=&-\frac{2}{9\pi^2}\frac{\alpha_{s}}{\pi }q^{2}\left [\log \left ( \frac{-q^{2}}{\nu ^{2}} \right )-\frac{3}{2}\right]\\
&+\frac{8}{9}\frac{\alpha _{s}}{\pi }\frac{1}{q^{2}}\left [\log \left ( \frac{-q^{2}}{\nu ^{2}} \right )-\frac{49}{6}  \right ]\left \langle m\bar{q}q \right \rangle\\
&-\frac{2}{3}\frac{\alpha _{s}}{\pi }\frac{1}{q^{2}}\left \langle G^{2} \right \rangle-\frac{16}{9}\frac{g^{2}}{q^{4}}\left \langle \bar{q}q\right \rangle^{2}+\frac{1}{72}\frac{g^{3}}{\pi ^{2}}\frac{1}{q^{4}}\left \langle G^{3} \right \rangle+\frac{8}{3}\frac{g}{q^{4}}\left \langle m\bar{q} Gq\right \rangle\textrm{,}
\end{aligned}
\end{equation}
where we use the BMHV scheme to calculate traces in $D$ dimension to keep
anti-commutativity of $\gamma_{5}$ \cite{ref_article13,ref_article14}.

By using the 50\%-10\% method described above, we obtain the sum rules window for $\tau$ in the range of
$( 0.32 \textrm{GeV}^{-2},0.62 \textrm{GeV}^{-2} )$  for the $1^{--}$ mixing correlator.
Minimizing $\chi^2$ leads to the solution
\begin{equation}
\left\{s_{0},\left|\frac{f_{1}f^{*}_{2}+f^{*}_{1}f_{2}}{2}\right|,m_{\bar{q}Gq\bar{q}q}\right\}=\left\{3.12^{+0.15}_{-0.13} \textrm{GeV}^{2}, 0.0126^{+0.0006}_{-0.0006} \textrm{GeV}^{4}, 0.737^{+0.058}_{-0.050} \textrm{GeV}\right\}.
\end{equation}

The decay constants of currents $j_{1\mu}$ and $j_{2\mu}$ with pure $1^{--}$ hybrid and $\bar qq$ states respectively  are $f_{\bar{q}q}^{'}=(0.770 \textrm{GeV})\times(0.153 \textrm{GeV})$
for the  pure $1^{--}$ $\bar{q}q$ state,  and $f_{\bar{q}Gq}^{'}=(2.34^3 \textrm{GeV}^3)\times(0.024 \textrm{GeV})$ for the pure $1^{--}$ hybrid state \cite{ref_article21,ref_article24}. We consistently absorb  mass factors in the definition of decay constants as described in the previous section.
The mixing strength can then be estimated by computing the value of $N$
\begin{equation}\label{firstn}
N_{\bar{q}Gq\bar{q}q}=\frac{0.0126\textrm{GeV}^4\times m_{\textrm{mix}}^2}{0.118\textrm{GeV}^2\times0.308\textrm{GeV}^4}=0.19\textrm{,}
\end{equation}
where $m_{\text{mix}}$ is the mixed state mass, i.e., $m_{\textrm{mix}}=m_{\bar{q}Gq\bar{q}q}$.
The result $N_{\bar{q}Gq\bar{q}q}$=0.19 shows that the mixing strength is not as weak as expected
since the mass of mixed state 0.770 GeV is very close to the $\rho$ meson, which  is usually considered
a very pure $\bar{q}q$ state . Then $N_{\bar{q}Gq\bar{q}q}$=0.19 is  the  strength (relative to the pure
hybrid meson) of the  $\rho$ meson coupling to the hybrid current. This suggests that the
$\rho$ meson contribution to the sum rules for the $1^{--}$ hybrid is negligible.  
From these results we find that the hybrid component of the
$\rho(1450)$ and $\rho(1570)$ is a few percent. However, we do not exclude a tetraquark and $\bar{q}q$ mixed state in this paper, which is a very
possible $1^{--}$ mixed state in the same mass range.

\section{Scalar  $\bar{q}q$ and glueball mixed state}\label{sec:IV}

In this section, we define
\begin{equation} \label{current}
\begin{aligned}
&j_{3u}=\frac{1}{2}m_q(\bar{u}(x)u(x)+\bar{d}(x)d(x))\textrm{,}\\
&j_{3s}=m_s\bar{s}(x)s(x),\\
&j_{4}=2\textrm{Tr}[\alpha_{s}G_{\mu\nu}(0)G_{\mu\nu}(0)],
\end{aligned}
\end{equation}
where $m_q$=$(m_u+m_d)/2$, to study the mixing between $0^{++}$ quark-antiquark state ($\bar uu+\bar dd$ or $\bar ss$) and glueball state.

Obviously chiral symmetry is broken if the off-diagonal mixing correlator
\begin{equation}
\Pi(q^{2})=i\int d^{4}x\, e^{iq\cdot x} \left \langle 0\left|j_{3u/3s}\left( x \right)j^{+}_{4}\left( 0 \right)  \right|0\right \rangle
\label{nondiag_corr}
\end{equation}
is non-zero, thus the perturbative contribution to the correlator must be proportional to the quark mass. As in Ref.~\cite{Harnett:2008cw}, we note that the effect of the
renormalization of the glueball current $j_{4}$ involving operator mixing must be included \cite{ref_article12}, i.e., we should use
the renormalized form of the glueball current
\begin{equation} \label{renormalization}
\left [ G_{\mu \nu }^{a}G_{\mu \nu }^{a} \right ]=\left ( 1-\frac{9}{2}\frac{\alpha _{s}}{\pi } \frac{1}{\epsilon }\right )G_{\mu \nu }^{aB}G_{\mu \nu }^{aB}+8\frac{\alpha _{s}}{\pi }\frac{1}{\epsilon }m^{B}\bar{q}^{B}q^{B}
\end{equation}
in our calculation, where the upper script $B$ denotes bare quantities, and we set $D=4-\epsilon$ in the  $\overline{\textrm{MS}}$ scheme.
As note in \cite{Harnett:2008cw}, the operator-mixing part of (\ref{renormalization}) will cancel $\log(-q^{2})/\epsilon$ divergence in the off-diagonal mixing correlator and keep
the imaginary part of the correlator finite. The following gluon propagator has been used in our calculation \cite{ref_article11}
\begin{equation}
\begin{aligned}
&\int dx \,e^{iq\cdot x} D^{ab}_{\mu \nu }\left ( x,y \right )=-\frac{g_{\mu \nu }}{q^{2}}e^{iqy}\delta_{ab}+\frac{2g}{q^{4}}G_{\mu \nu }^{ab}\left ( 0 \right )e^{iqy}-giy_{\varphi }G_{\varphi \rho }^{ab}\left ( 0 \right )\frac{q_{\rho }}{q^{4}}e^{iqy}g_{\mu \nu }\textrm{,}\\
\end{aligned}
\end{equation}
where $G_{\mu \nu }^{ab}=f^{abc}G_{\mu \nu }^{c}$.

\begin{figure}[ht]
 \centering \includegraphics[width=0.8\columnwidth]{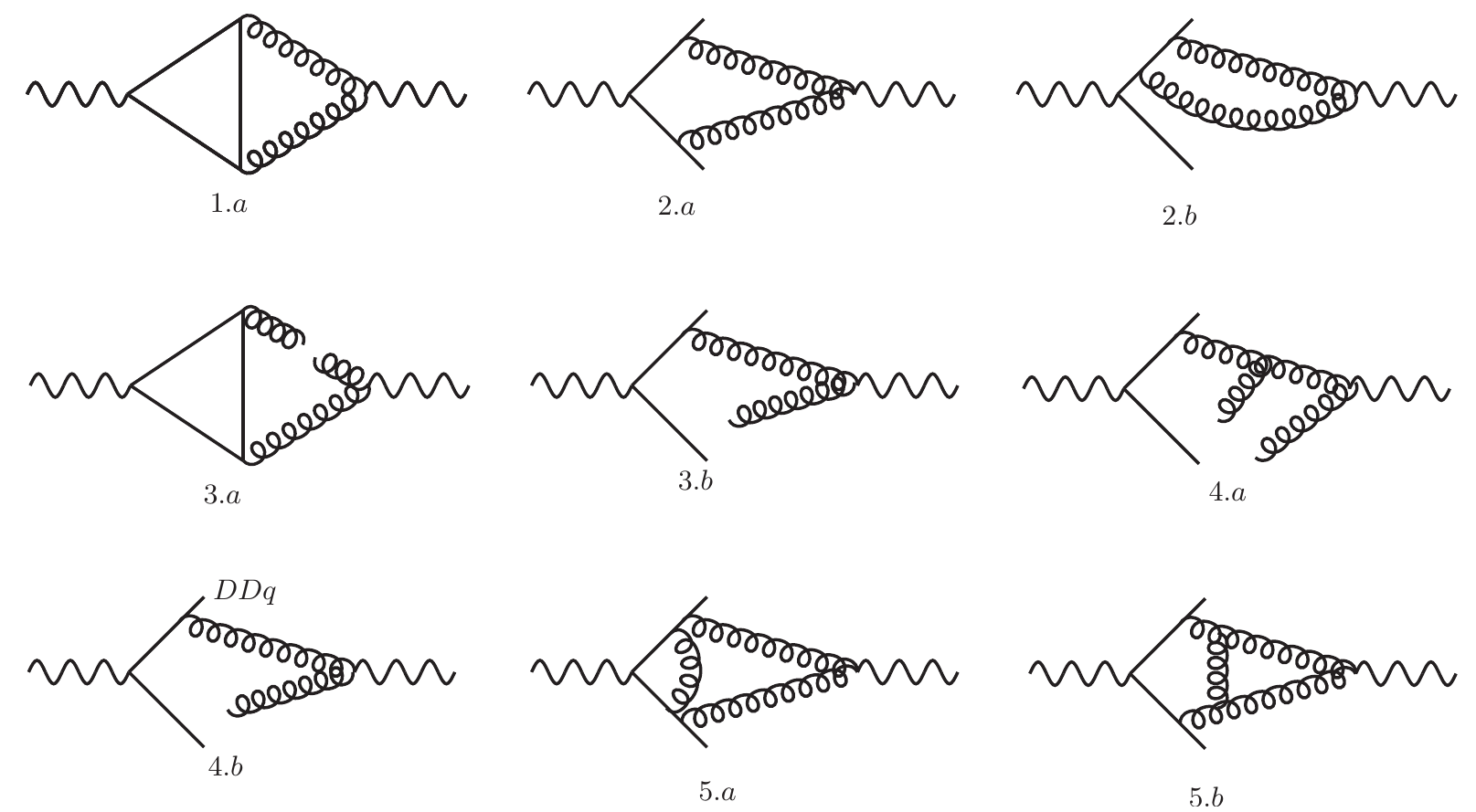}
  \caption{
                \label{fig6} 
              Feynman diagrams for the $\bar{q}q$-glueball mixed state. We omit most of the radiative correction diagrams which have  been computed, and just list two of them at the end of the diagram terms. Note that 4.b just lists one expansion form.
        }
\end{figure}

The OPE result for the $\bar{q}q$-glueball mixed correlator  from Figure \ref{fig6} is
\begin{equation} \label{qqgg}
\begin{aligned}
\Pi_{\bar{q}qGG}(q^{2})=&-m^{2}q^{2}\log \left ( -\frac{q^{2}}{\nu^2} \right )\left [ -\frac{23}{2\pi} \left ( \frac{\alpha_s }{\pi } \right )^{2}+\frac{3}{2\pi }\left ( \frac{\alpha_s }{\pi } \right )^{2}\log \left ( -\frac{q^{2}}{\nu^2} \right )\right ]\\&+\left [ -8\pi \left ( \frac{\alpha_s }{\pi } \right )^{2} +18\pi \left ( \frac{\alpha_s }{\pi } \right )^{3}\log \left ( -\frac{q^{2}}{\nu^2} \right ) -\frac{326}{3}\pi \left ( \frac{\alpha_s }{\pi } \right )^{3}\right ]\log \left ( -\frac{q^{2}}{\nu^2} \right )\left \langle m\bar{q}q \right \rangle\\&-\frac{m^{2}}{q^{2}}\left [ 6  \frac{\alpha_s }{\pi }  -2 \frac{\alpha_s }{\pi } \log \left (-\frac{q^{2}}{\nu^2} \right )  \right ]\left \langle \alpha_s G^2 \right \rangle-4\frac{\alpha_s }{q^{2}}\left \langle gm\bar{q}Gq \right \rangle\\&+3 \frac{\alpha_s }{\pi }  \frac{A}{q^4}\left \langle gm\bar{q}Gq \right \rangle+\frac{4}{3}\frac{\alpha_s\pi  }{q^{4}}\left \langle m\bar{q}q \right \rangle\left \langle \alpha_s G^2 \right \rangle \textrm{,}
\end{aligned}
\end{equation}
where 8 dimension operators are factorized in order to conduct a QCDSR analysis
\begin{equation}
\begin{aligned}
&\left \langle m\bar{q}G_{\mu \nu }G_{\mu \nu } q\right \rangle=\left \langle m\bar{q}q \right \rangle\left \langle G^2 \right \rangle\textrm{,}\\
&\left \langle m\bar{q}\left [ G_{\mu \lambda },G_{\nu \lambda }\right ]\sigma _{\mu \nu } q\right \rangle=A\left \langle m\bar{q}G q \right \rangle\textrm{,}
\end{aligned}
\end{equation}
with $A=\frac{g^{3}\left \langle G^{3} \right \rangle}{\alpha_{s}\left \langle G^{2} \right \rangle}$.
In Eq.~(\ref{qqgg}), $\bar{q}q$=$\frac{1}{2}(\bar{u}u+\bar{d}d)$, $m=m_q$ for $u,d$ quark case and $\bar{q}q$=$\bar{s}s$, $m=m_s$ for $s$ quark case. Our calculation confirms the perturbative, gluon condensate, mixed condensate, and leading-order   quark  condensate  of Ref.~\cite{Harnett:2008cw} and extends the results for the correlator to include higher-dimension condensates and next-to-leading order quark condensate terms. We do not compute the radiative correction to the perturbative contribution because it is chirally suppressed and numerically small compared with the $\left \langle m\bar{q}q \right \rangle$ condensate.

\begin{figure}[ht]
 \centering \includegraphics[width=0.6\columnwidth]{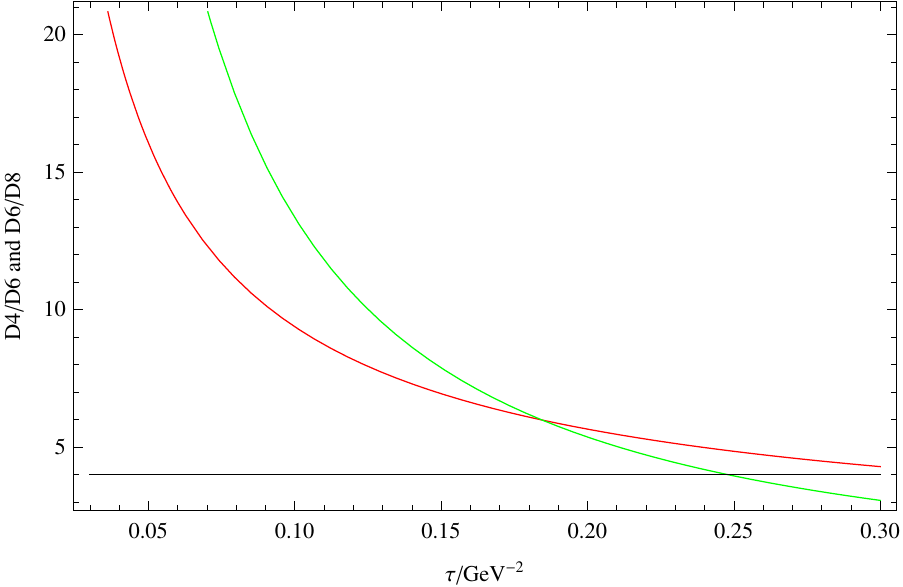}
  \caption{
                \label{fig3} 
       Ratios of different dimension contributions to the OPE are shown as a function of $\tau$ for  the  $u,d$ quark case.  The red curve represents the 4D/6D ratio and the green line corresponds to 6D/8D.
        }
        \label{ratio_qq}
\end{figure}

\begin{figure}[ht]
 \centering \includegraphics[width=0.6\columnwidth]{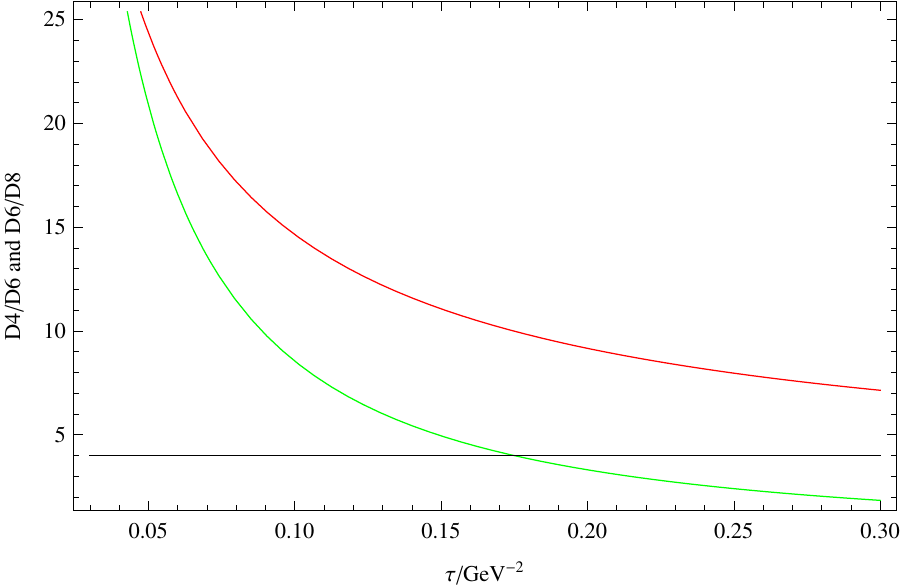}
  \caption{
                \label{fig7} 
              Ratios of different dimension contributions to the OPE are shown as a function of $\tau$ for  the  $s$ quark case.  The red curve represents the 4D/6D ratio and the green line corresponds to 6D/8D.       
        }
              \label{ratio_ss}
\end{figure}

The sum rule window, $\tau\in$ ( 0.03 GeV$^{-2}$, 0.25 GeV$^{-2}$) for the $u, d$ quark case, and
$\tau\in$ ( 0.08 GeV$^{-2}$, 0.17 GeV$^{-2}$) for the $s$ quark case are obtained by demanding that 8D contributions are 1/4 of the 6D, thus 8D contributions is less than 5\% of $R^{\textrm{(OPE)}}(\tau)$. With the presence of more terms in the OPE for the glueball mixing correlator, we are able to extend the criterion used for the hybrid to encompass OPE convergence from   higher-dimension condensates.  Ratios of different dimensional OPE contributions are shown in Figs.~\ref{ratio_qq} and \ref{ratio_ss}.
The solutions that minimize $\chi^2$ are
\begin{equation}\label{eq:result}
\begin{aligned}
&\left\{s_{0},\left|\frac{f_{3u}f^{*}_{G}+f^{*}_{3u}f_{G}}{2}\right|,m_{\bar{u}uGG}\right\}=\left\{18.13^{+0.42}_{-0.42} \textrm{GeV}^{2}, 0.000291^{+0.000032}_{-0.000029} \textrm{GeV}^{6}, 0.867^{+0.052}_{-0.066} \textrm{GeV}\right\}\textrm{,}\\
&\left\{s_{0},\left|\frac{f_{3s}f^{*}_{G}+f^{*}_{3s}f_{G}}{2}\right|,m_{\bar{s}sGG}\right\}=\left\{15.21^{+0.26}_{-0.28} \textrm{GeV}^{2}, 0.0158^{+0.0027}_{-0.0022} \textrm{GeV}^{6}, 1.654^{+0.076}_{-0.085} \textrm{GeV}\right\}\textrm{,}
\end{aligned}
\end{equation}
where the first line is solution for $u,d$ quark case, and the second line is solution for $s$ quark case.
The result (\ref{eq:result}) suggests that $f_0(500)$ and $f_0(1710)$ are respectively good candidates for mixed states of glueballs with $\bar qq$ and $\bar s s$ components. The $f_0(980)$ may also couple to $\bar qq$, but its coupling should be much weaker than the $f_0(500)$ \cite{ref_article27}.
There is excellent agreement between the $\bar q q$ mass prediction \eqref{eq:result} and that of Ref.~\cite{Harnett:2008cw}.  Taking into account the differences from Ref.~\cite{Harnett:2008cw} in the  field-theoretical content (e.g., higher-dimension condensates) and the use of Laplace versus Gaussian sum-rules that weight the OPE terms in fundamentally different ways, the agreement is remarkable. 
The $\bar s s$ case was not studied in Ref.~\cite{Harnett:2008cw}  so no comparison with previous work is possible.

Finally, we estimate the mixing strength of these two states with same method we have used in the $1^{--}$ case.
The relevant decay constants for the pure $0^{++}$ $\bar qq$ state and glueball have been obtained in literature:
$f^{'}_{q}$=0.5GeV$\times$0.64GeV for $\frac{1}{\sqrt{2}}(\bar{u}u+\bar{d}d)$, $f^{'}_{s}$=0.98GeV$\times$0.41GeV for
$\bar{s}s$, $f^{'}_{G}$=$1.53^2$GeV$^2$$\times$1.01GeV for glueball \cite{ref_article27,ref_article28,ref_article29}.
Collecting all the parameters we compute $N_{\bar{u}uGG/\bar{s}sGG}$
\begin{equation}
\begin{aligned} N_{\bar{u}uGG}=\frac{0.000291\textrm{GeV}^6}{0.32\textrm{GeV}^2\times2.36\textrm{GeV}^3\times m_{q}}=0.11\textrm{,}\\
N_{\bar{s}sGG}=\frac{0.0158\textrm{GeV}^6}{0.40\textrm{GeV}^2\times2.36\textrm{GeV}^3\times m_{s}}=0.17\textrm{,}
\end{aligned}
\end{equation}
where we have used masses of $u$, $d$ and $s$ quarks at the energy scale $\mu_{0}$=2 GeV\cite{ref_article1}, which corresponds to the energy scale at which decay constants were derived in \cite{ref_article27}(since $m_{q}f^{'}_q$ and $m_{s}f^{'}_s$ are energy scale independent), i.e.,
\begin{equation}
m_{q}=\left(m_u+m_d\right)/2=3.5~\textrm{MeV}, ~m_{s}=96~\textrm{MeV}.
\end{equation}
We divide by the quark mass in the above calculation because of different  $\bar{q}q$ current definitions.

Comparing to $N$ in $1^{--}$, we see that the mixing strength of these two states are
smaller but close to the $1^{--}$ case. It shows that these states are weakly coupled with one of
two currents. This result should not be surprising because the mixing is chirally suppressed
in the perturbative  corrections. The mixing is dominated by non-perturbative chiral-violating condensate contributions, which converge slowly. In order to make the OPE series convergent, we have to use a window with a large energy scale where the low-energy condensate terms are suppressed. The coupling we obtain
is therefore referenced to an energy scale far from the mass of the state, so our conclusion may  change when
the energy scale  is decreased  to the resonance mass. Furthermore, there is a large mass difference
between  $f_0(500)$ and $f_0(1710)$, which cannot be explained by the difference in the
quark masses and condensates, so we suggest that, for the relative mixture of $\bar q q$ and gluonic content, the $f_0(500)$ is dominated by a $\bar{u}u+\bar dd$ component and  $f_0(1710)$ is
dominated by a  glueball component.  It is important to emphasize that the off-diagonal correlator \eqref{nondiag_corr} explores  the mixing of gluonic and $\bar qq$ components, so our analysis does not constrain tetraquark components of the $f_0$ states.  

A meaningful comparison with the mixing results of Ref.~\cite{Harnett:2008cw} is difficult because different aspects were explored.  In \cite{Harnett:2008cw}, the Gaussian sum-rule permitted study of multiple states in the off-diagonal correlator,  and a strong mixing between these states was found, but no analysis of pure states was performed to allow comparison with the mixing parameter $N$.  Nevertheless,  Ref.~\cite{Harnett:2008cw} does also conclude that an approximately $1.5\,{\rm GeV}$ state is predominantly  a glueball.  

Because of the proximity in mass, it is easy to understand that there is a stronger mixing between
$\bar ss$ and glueball than $\bar uu+\bar dd$ and glueball, however, it is subtle why the
sum rules select a heavier glueball mixing with a few percent of  $\bar ss$ rather than a
lighter $\bar ss$ with a few percent of glueball.  The only reason  emerging from the sum rules
is that the couplings of $\bar ss$ state to the currents (\ref{current}) is not as strong as
those of the glueball state. 

 In principle, it is possible that there is more than one state
giving comparable contributions  to the correlator  (even though the simulation shows single pole
model works well), then the average mass of these contributing states should be between 600 MeV and 1700 MeV
(i.e., the range of the low lying  $0^{++}$ states found in experiments). We could use the
mass prediction in Eq.~\eqref{eq:result} to estimate the mixing information. In the $u,d$ quark case, the average mass
is 867 MeV. This excludes the case that heavier states have a large contribution to the mixing. In the $s$ quark case,
the result also prohibits a large contribution from the states much lighter than 1600 MeV.

Superficially one might expect a large sensitivity to the quark mass parameters  because the perturbative process in the mixing correlator are proportional to the square of the quark mass.  However, the constraints on the sum-rule working window of $\tau$ limit the sensitivity to these perturbative corrections.  As the lower limit on $\tau$ is decreased, the  single pole model  will begin to fail  (i.e., the excited states and continuum  will give a large 
contribution) while the perturbative contributions for non-strange quarks  remains small.    As the upper limit on $\tau$ is increased  to 
$\tau=0.3~\textrm{GeV}^{2}$ OPE convergence begins to fail.  Thus for any reasonable variation in the working window in $\tau$  we still find that non-perturbative contributions will dominate the mixing.

\section{Summary}

In this paper we have used QCD sum-rule methods  for off-diagonal correlation functions to  study  the mixing of $\bar q q$ with hybrid and glueball components for the $1^{--}$ vector and $0^{++}$  scalar channels.
 The mass prediction for the $1^{--}$  mixed state is $0.737^{+0.058}_{-0.050}$ GeV, consistent with the mass of $\rho$(770) within the errors and very close to the
mass obtained  in QCDSR using solely the vector interpolating current $\bar q \gamma_\mu q$
\cite{ref_article24}. This result disfavours a  large mixing between $1^{--}$ light $\bar qq$
and hybrid mesons.

For $0^{++}$ particles,  we find the mass predictions are respectively $0.867^{+0.052}_{-0.066}$
GeV for the $\bar{u}u+\bar{d}d$-$GG$ mixed state and $1.654^{+0.076}_{-0.085}$ GeV for the
$\bar{s}s$-$GG$ mixed state. These results qualitatively show that $f_0(500)$ and $f_0(1710)$
can be candidates in these two cases. We have estimated the ``mixing strength'' defined in
\eqref{eq:mixing dgree} for these states which represents whether the $\bar q q$ and gluonic components of the  ``physical state''
under consideration is more of a mixed state or a pure state. From the mixing strength one
can see that the $\bar q q$ and gluonic components of these scalar states are likely not to be strongly mixed, with the $f_0(1710)$ being close to a pure glueball.
 In fact, $\rho$(770) is generally considered
as a very pure state (which can also be seen from our previous analysis). If we set the $\rho$ meson as a
standard to examine other states, we see that $f_0(500)$ is even more pure than $\rho$(770),
while $f_0(1710)$, which was considered as a strongly mixed state of a $\bar{q}q$ meson and a
glueball, has a similar mixing strength as $\rho$(770).

As noted earlier, our analysis through the off-diagonal correlator \eqref{nondiag_corr} examines the mixing of  the $\bar q q $ and gluonic components, so our results do not  constrain the four-quark content of the $f_0(500)$.  However, we can conclude that the $f_0(500)$ has both gluonic and $\bar q q$ components because this state emerges from the mixed correlator \eqref{nondiag_corr}, and our small-mixing result suggests that the relative proportion of one of  component may dominate the other.  In some previous analyses the relative proportion of gluonic components are more prominent than $\bar q q$ (see e.g.,  
\cite{Narison:1996fm,Narison:2005wc,Ochs:2013gi,Mennessier:2010xg,Minkowski:1998mf}) , while in other approaches it is the opposite  (see e.g., \cite{06_F,Albaladejo:2008qa}).

Finally, it should be noted that the accuracy of our work is subject to some factors. For the
$0^{++}$ case, we are constrained  to do our analysis in a window with a relatively large Borel scale
$\frac{1}{\tau}$ (compared to the resonance mass) to ensure  OPE convergence, which however may also suppress the non-perturbative QCD effects that dominate the mixing. Furthermore, our
analysis is also sensitive to the decay constants of the pure states (such as the $1^{--}$
hybrid and the $0^{++}$ glueball) which have not been measured in experiments and thus rely
on sum rule determinations and model calculations that may have input parameters in tension with our work. 

Clearly it is complicated to rigorously
consider the mixing in QCD, and our sum rule analysis provides estimates which suggest the
effects of the mixing between hybrids/glueballs and ordinary $\bar qq$ mesons are very limited
in the vector and scalar channels. The methods of this work can be extended to the mixing between
tetraquarks and hybrids or $\bar qq$ states. It would be of particular interest if such mixing
is not be suppressed in the sum-rule working interval by some (approximate) symmetries such as the chiral symmetry.

\acknowledgments
This work is supported by NSFC (under grants 11175153 and 11205093) and the China Postdoctoral Science
Foundation funded project (2018M631572). TGS is grateful for financial support from the Natural Science and Engineering Research Council of Canada (NSERC).

\appendix
\section{QCDSR fitting results}
\begin{figure}[ht]
 \centering \includegraphics[width=0.6\columnwidth]{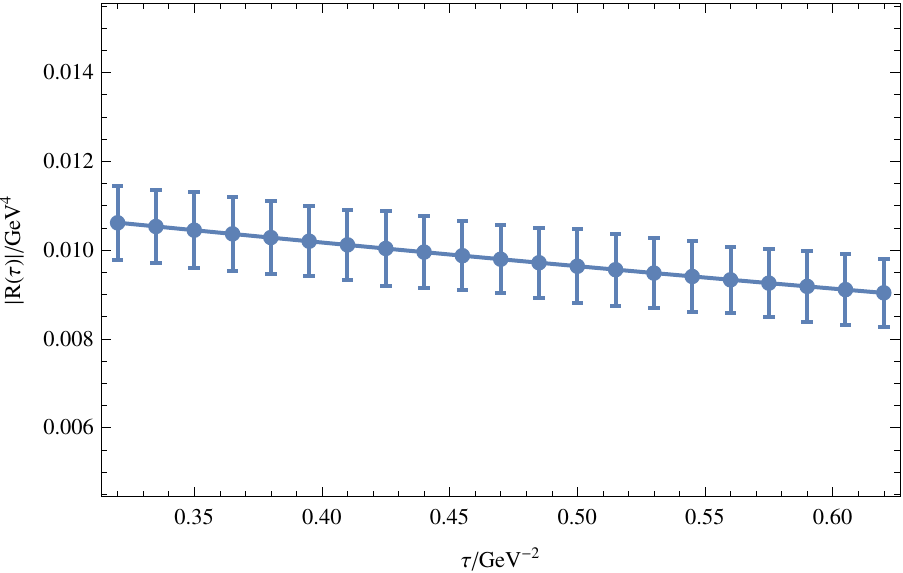}
  \caption{
                \label{fig8} 
              QCDSR minimum-$\chi^2$ fit for the $\bar{q}q$-hybrid mixed state: the dots represent the QCD side of the master equation (\ref{2}), and the middle line represents phenomenological side of (\ref{2}). The error bars are induced by 10\% uncertainty of the phenomenological parameters  of Table~\ref{tbl:bins}.
        }
\end{figure}
\begin{figure}[ht]
 \centering \includegraphics[width=0.6\columnwidth]{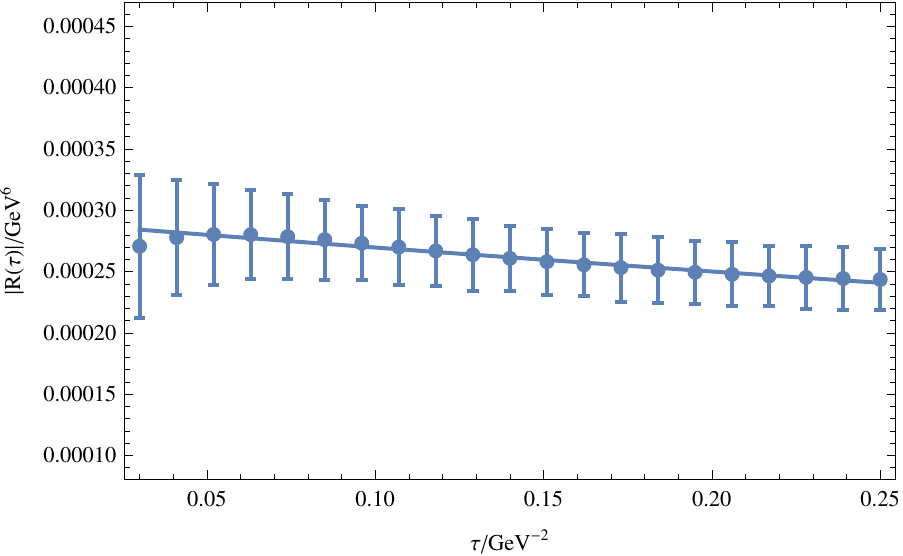}
  \caption{
                \label{fig9} 
               QCDSR minimum-$\chi^2$ fit 
             for the $\bar{u}u$+$\bar{d}d$ and glueball mixed state: the dots represent QCD side of the master equation (\ref{2}), and the middle line represents phenomenological side of (\ref{2}).  The error bars are induced by 10\% uncertainty of the phenomenological parameters in of the phenomenological parameters  of Table~\ref{tbl:bins}.
        }
\end{figure}\begin{figure}[ht]
 \centering \includegraphics[width=0.6\columnwidth]{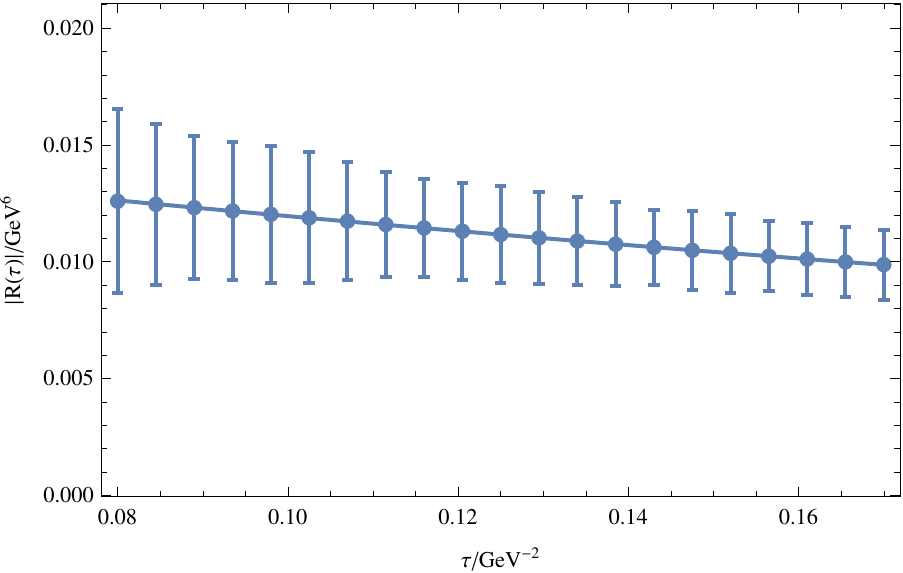}
  \caption{
                \label{fig10} 
                QCDSR minimum-$\chi^2$ fit 
             for the $ss$ and glueball mixed state: the dots represent QCD side of the master equation (\ref{2}), and the middle line represents phenomenological side of (\ref{2}).  The error bars are induced by 10\% uncertainty of the phenomenological parameters in of the phenomenological parameters  of Table~\ref{tbl:bins}
        }
\end{figure}

\end{document}